\documentclass[11pt,onecolumn,english,onecolumn]{IEEEtran}
\usepackage[T1]{fontenc}
\usepackage[latin9]{inputenc}
\usepackage{float}
\usepackage{amsmath}
\usepackage{amsthm}
\usepackage{amssymb}
\usepackage{geometry}
\usepackage{setspace}
\makeatletter

\floatstyle{ruled}
\newfloat{algorithm}{tbp}{loa}
\providecommand{\algorithmname}{Algorithm}
\floatname{algorithm}{\protect\algorithmname}

\theoremstyle{plain}
\newtheorem{thm}{\protect\theoremname}
\theoremstyle{plain}
\newtheorem{prop}[thm]{\protect\propositionname}
\theoremstyle{plain}
\newtheorem{cor}[thm]{\protect\corollaryname}
\theoremstyle{remark}
\newtheorem{claim}[thm]{\protect\claimname}
\theoremstyle{plain}
\newtheorem{lem}[thm]{\protect\lemmaname}

\usepackage{hyperref}
\usepackage{enumitem}
\usepackage{breqn}
\usepackage{bbm} 
\usepackage{cite}
\usepackage{algorithm,algpseudocode}

 \DeclareMathOperator*{\argmin}{arg\,min}

\global\long\def\P{\mathbb{P}}
\global\long\def\E{\mathbb{E}}

\global\long\def\I{\mathbbm{1}}
\global\long\def\d{\mathrm{d}}

\global\long\def\lcb{\mathrm{LCB}}

\global\long\def\pregret{\mathsf{preg}}

\global\long\def\trre[#1,#2]{\overset{{\scriptstyle (#2)}}{#1}} % transition explained with reason
\renewcommand\[{\begin{equation}}
\renewcommand\]{\end{equation}}

\allowdisplaybreaks

\author{Nir Weinberger$^{1}$ and Ram Zamir$^{2}$}

\makeatother

\usepackage{babel}
\providecommand{\claimname}{Claim}
\providecommand{\corollaryname}{Corollary}
\providecommand{\lemmaname}{Lemma}
\providecommand{\propositionname}{Proposition}
\providecommand{\theoremname}{Theorem}

\begin{document}
\title{Exploration-Exploitation Tradeoff in Universal Lossy Compression\thanks{$^{1}$ Nir Weinberger is with the Viterbi Faculty of Electrical and
Computer Engineering, Technion -- Israel Institute of Technology,
Haifa 3200004, Israel (Email: nirwein@technion.ac.il.). $^{2}$ Ram
Zamir is with the Department of Electrical Engineering-Systems, Tel
Aviv University, Ramat Aviv, 6997801, Israel (Email: zamir@eng.tau.ac.il).
This paper was presented in part in the 2025 IEEE International Symposium
on Information Theory, June 2025, Ann Arbor, MI, US. The research
of N. W. was supported in part by the Israel Science Foundation (ISF),
grant no. 1782/22.}}

\maketitle
\renewenvironment{align*}{\align}{\endalign}
\thispagestyle{empty}
\begin{abstract}
Universal compression can learn the source and adapt to it either
in a batch mode (forward adaptation), or in a sequential mode (backward
adaptation). We recast the sequential mode as a multi-armed bandit
problem, a fundamental model in reinforcement-learning, and study
the trade-off between exploration and exploitation in the lossy compression
case. We show that a previously proposed ``natural type selection''
scheme can be cast as a reconstruction-directed MAB algorithm, for
sequential lossy compression, and explain its limitations in terms
of robustness and short-block performance. We then derive and analyze
robust cost-directed MAB algorithms, which work at any block length.
\end{abstract}

\section{Introduction}

\subsection{Exploration-versus-Exploitation}

Universal data compression can work either in a \emph{batch mode},
which typically results a large delay, or in a \emph{sequential, on-line,
mode}, which operates instantaneously, and typically needs a much
smaller delay. The sequential mode is based on backward adaptation,
that is, on the the knowledge of the data source statistics that is
common to both the encoder and decoder. Backward adaptation is fundamentally
more challenging in the case of lossy compression than in lossless
compression \cite{davisson1973universal,krichevsky1981performance,rissanen1984universal,LZ1,LZ2,ziv1978compression,shtar1987universal,CTW1995},
because the decoder only observes the noisy (quantized) past reconstructions.
From a different perspective, the decoder only receives a partial
feedback regarding the quality of its past decisions. Thus, an exploration-exploitation
trade-off is exhibited, as in other similar sequential decision-making
problems, such as the multi-armed bandits (MAB) \cite{bubeck2012regret,slivkins2019introduction,lattimore2020bandit},
and more generally, \emph{reinforcement learning} (used, in the context
of lossy compression, e.g., in \cite{cregg2024reinforcement}). Our
first step is therefore to propose a unified online learning framework,
which allows to discuss under the same footing, both classic, expert-based,
algorithms, as well as data-driven algorithms such as MAB-based ones.

\subsection{A Unified Online Learning Framework \label{sec:A-Unified-Online}}

Consider the following unified framework for sequential lossy compression.
Let $\{X_{t}\}_{t\in\mathbb{N}_{+}}$ be an information source from
an alphabet ${\cal X}$. The abstract alphabet ${\cal X}$ can represent
either a scalar or a vector, and can be either discrete or continues.
We focus on the stochastic setting, in which the source is memoryless
and $X_{t}\sim P_{X}$, independently and identically distribution
(IID) though this distribution is unknown in advance (\emph{a universal
setting}). The source sample $X_{t}$ at the encoder side is reconstructed
as $Y_{t}$ from an alphabet ${\cal Y}$ at the decoder side. A distortion
measure is given by $\rho:{\cal X}\times{\cal Y}\to\mathbb{R}$, as
well as a target distortion level $d$. Before the compression of
a source symbol $X_{t}$, the encoder and decoder agree on a given
codebook ${\cal C}_{t}\subset{\cal Y}$. We assume, for simplicity,
that for each input symbol $X_{t}\in{\cal X}$ there exists $Y_{t}\in{\cal C}_{t}$
such that $\rho(X_{t},Y_{t})\le d$. The encoder uses a variable-length
prefix-free code to send $Y_{t}$ to the decoder.

Concretely, the codebook ${\cal C}_{t}$ at each time point $t\in\mathbb{N}_{+}$
is generated as follows. The encoder and decoder agree on taking an
action $A_{t}$ from a set ${\cal A}$. The action determines, in
some pre-specified way, a codebook that is agreed by the encoder and
decoder. In this paper, we focus on the case in which the action determines
a reconstruction distribution $Q_{t}=f(A_{t})$, where $f\colon{\cal A}\to{\cal P}({\cal Y})$
is some rule. Accordingly, let ${\cal Q}=\{f(a)\}_{a\in{\cal A}}$
be the set of possible reconstruction distributions. Using shared
randomness, the encoder and decoder both generate a random codebook
${\cal C}_{t}=\{Y_{t}(1),Y_{t}(2),\ldots,\}\subseteq{\cal Y}$ (in
principle, of infinite length), where $Y_{t}(j)\sim Q_{t}$, are drawn
in an IID manner. For a given $x\in{\cal X}$ and a codebook ${\cal C}=\{y(1),y(2),\ldots\}$,
let
\[
J(x,{\cal C}):=\inf_{j\in\mathbb{N}_{+}}\left\{ j\colon\rho\left(x,y(j)\right)\leq d\right\} 
\]
be the first \emph{$d$-match}, i.e., the minimal codeword index satisfying
the distortion constraint $d$, and $J(x,{\cal C})=\infty$ if no
such codeword exists in ${\cal C}$. Then,
\[
b(x,{\cal C}):=\log_{2}J(x,{\cal C})
\]
is the number of bits required for encoding $x$ to the decoder using
the codebook ${\cal C}$. Thus, upon observing the sample $X_{t}$,
the encoder finds $J(X_{t},{\cal C}_{t})$ and encodes it using $B_{t}:=b(X_{t},{\cal C}_{t})$
bits. For a given $x\in{\cal X}$ and reconstruction distribution
$Q\in{\cal P}({\cal Y})$, the average number of bits required for
encoding at distortion $d$ is thus 
\[
\overline{b}(x,Q)=\E_{{\cal C}\sim^{\text{IID}}Q}\left[b(x,{\cal C})\right],
\]
i.e., averaged over the random choice of the codebook.

To allow the encoder and decoder synchronize the choice of action
$A_{t}$, and thus the distribution $Q_{t}$, the choice is based
only on the information that is known to both sides, i.e., on\emph{
backward adaptation.} The decoder does not observe the clean samples
$X_{t}$, but only their reconstruction $Y_{t}$, and so the common
\emph{history} at time $t\in\mathbb{N}_{+}$ is 
\[
{\cal H}_{t}:=(A_{1},B_{1},Y_{1},\ldots A_{t-1},B_{t-1},Y_{t-1}).
\]
The policy $\pi_{t}$ that determines $A_{t}=\pi_{t}({\cal H}_{t})$
where 
\[
\pi_{t}\colon({\cal P}({\cal Y})\otimes\mathbb{\mathbb{R}}_{+}\otimes{\cal Y})^{\otimes(t-1)}\to{\cal A}.
\]

The setup above is an online learning game \cite{cesa2006prediction}
in which the decoder is the agent whose action $A_{t}$ is the choice
of distribution $Q_{t}$, based on the past actions and observations
${\cal H}_{t}$. The generator of the $\{X_{t}\}$ combined with the
encoder are the Nature player, which generates a random \emph{cost}
$b(X_{t},{\cal C}_{t})$ based on the choice of $A_{t}$. Given an
action, the randomness of the cost stems from both the randomness
of the source sample $X_{t}$ and the codebook ${\cal C}_{t}$. The
agent (decoder) only has partial feedback on the loss of other actions,
as it cannot observe the loss of other reconstruction distributions.
Thus, this is a \emph{partial monitoring} problem \cite[Ch. 37]{lattimore2020bandit}. 

Given the set of reconstructions ${\cal Q}$, a source-optimal encoder
compressed the sources at a minimal cost (expected rate) of 
\begin{equation}
R^{*}:=\min_{Q\in{\cal Q}}\E_{X\sim P_{X}}\left[\overline{b}(X,Q)\right].\label{eq: optimal cost}
\end{equation}
Hence, the \emph{pseudo-regret} \cite[Ch. 1]{slivkins2019introduction},
\cite[Ch. 1]{bubeck2012regret} of the policy $\pi_{t}$ at time $t\in\mathbb{N}_{+}$
is
\begin{equation}
\pregret_{t}:=\sum_{i=1}^{t}\E_{X\sim P_{X}}\left[\overline{b}(X,Q_{i})\right]-R^{*}.\label{eq: pseudo regret}
\end{equation}

The goal of an online learner is to choose the policy $\{\pi_{t}\}$
in order to minimize the pseudo-regret. The trade-off between \emph{exploitation
---} choose $Q_{t}$ to be the minimal cost reconstruction from the
reconstruction distributions attempted thus far, and \emph{exploration}
--- choose $Q_{t}$ from an action that has not been played before,
which is potentially better. The optimal balance between these, possibly
conflicting, goals is the subject of a well-studied MAB framework
\cite{bubeck2012regret,slivkins2019introduction,lattimore2020bandit}.

By contrast, in the realm of lossy compression, an expert-based approach
for backward-adaptive lossy compression, called \emph{natural type
selection} (NTS) \cite{NTS_original,ZamirRose1997}, was developed.
In the NTS scheme, the decoder chooses the next reconstruction distribution
based on the previous reconstruction $Y_{t-1}$ (specifically, its
\emph{type}). The NTS scheme was extended in various forms including
parametric models \cite{kochman2003computation,NTS_parametric}, memory
models \cite{NTS_memory} incremental implementation \cite{NTS_ISIT},
to channel coding \cite{TridenskiZamir2020}, continuous alphabets
\cite{NTS_abstract_alphabet}, non-convex sets \cite{Hila-Rami2023}.
However, they are restricted to the asymptotic regime, where large-deviation
principles apply, with the recent \cite{NTS_ISIT} being an exception;
see below. 

\subsection{Contributions}

Since the NTS approach has been widely studied, we begin by reviewing
the NTS algorithm in light of the unified framework suggested above.
We then discuss its limitations, and the regime in which MAB-based
solutions are of interest. MAB algorithms were applied in various
information-theoretic coding setups (e.g., the recent \cite{weinberger2023multi,egger2024maximal}),
but, to the best of our knowledge, the first proposal to study a sequential
lossy compression version of it is in a precursor of the current paper
\cite{zamir2024alternate}. We thus consider basic MAB algorithms,
and specifically, those whose policy is cost-directed, that is, those
which base their policy only on the cost it incurred in previous rounds,
but not on the reconstruction. We develop two MAB algorithms, one
which is suitable for a given finite set of actions, yet allows for
an arbitrary reconstruction alphabet, and one which assumes a finite
reconstruction alphabet, yet allows to freely optimize the set of
actions. We use the standard upper confidence bound (UCB) algorithm
\cite[Sec. 2.2]{bubeck2012regret} \cite[Ch. 1]{slivkins2019introduction}
for the first one, and a Lipschitz-bandit \cite[Ch. 4]{slivkins2019introduction}
for the second one. We highlight the specific technical challenges
in adapting these algorithms and their analysis to the sequential
lossy compression, specifically including obtaining tight confidence
intervals for the cost of each action, and establishing the required
Lipschitz properties of the cost function. We characterize complexity
terms that affect the regret. We then conclude the paper with a discussion
on the potential to refine the MAB approach, and incorporate the reconstruction-directed
principles of the NTS into MAB algorithms. 

\section{Natural Type Selection}

The basic NTS algorithm addresses the standard lossy compression setting
\cite{Berger_RD}, where both $x$ and $y$ are length-$\ell$ vectors,
and the source distribution is memoryless. Specifically, for some
finite alphabets ${\cal U}$ and ${\cal V}$, it holds that ${\cal X}={\cal U}^{\otimes\ell}$
and ${\cal Y}={\cal V}^{\otimes\ell}$, and the distortion measure
is additive, that is, for $X_{t}=(X_{t}(1),\ldots,X_{t}(\ell))\in{\cal U}^{\otimes\ell}$
and $Y_{t}=(Y_{t}(1),\ldots,Y_{t}(\ell))\in{\cal V}^{\otimes\ell}$
it holds that $\rho(X,Y)=\sum_{i=1}^{\ell}\varrho(X(i),Y(i))$ for
some symbol-wise distortion measure $\varrho:{\cal U}\times{\cal V}\to\mathbb{R}_{+}$.
Moreover, for a random $X_{t}=(X_{t}(1),\ldots X_{t}(\ell))$ it holds
that $X_{t}(j)\sim P_{U}$, IID. In this case, the asymptotically
optimal distribution in terms of rate-distortion function $R(P_{U},d)$,
is a \emph{memoryless distribution} over ${\cal V}^{\otimes\ell}$
(that is, each codeword has IID symbols, and different codewords are
also IID) . Let ${\cal P}({\cal Z})$ be the set of probability distributions
over an alphabet ${\cal Z}$. That is, the optimal $Q^{*}\in{\cal P}({\cal Y})$
is such that $Q^{*}(y)=\prod_{i=1}^{\ell}Q_{V}^{*}(y_{i})$ for some
$Q_{V}^{*}\equiv Q_{V}^{*}(P_{U},d)\in{\cal P}({\cal V})$. 

We describe three types of NTS algorithms, arranged according to their
reliance on the asymptotic properties of large $\ell$, with the motivation
to back off from asymptotic behavior towards non-asymptotic setting
and MAB algorithms. First, we begin with a degenerate NTS algorithm
\cite{ZamirRose1997}, which is based on universal, mixture of types,
distributions. Second, we move to the an NTS algorithm \cite{NTS_original}
that operates on memoryless distributions. Third, we consider a recent
version \cite{NTS_ISIT}, which aims to be effective for non-asymptotic
blocklengths by averaging. Before that, let us briefly recall and
introduce various definitions related to type classes. 

\paragraph*{Distributions over Types}

Let ${\cal P}_{\ell}({\cal V})\subset{\cal P}({\cal V})$ be the set
of types (empirical distributions) for length-$\ell$ vectors over
${\cal V}$, that is,
\[
{\cal P}_{\ell}({\cal V}):=\left\{ P\in{\cal P}({\cal V})\colon\ell P(v)\in\mathbb{N}_{+}\text{ for all }v\in{\cal V}\right\} .
\]
The type class of $P\in{\cal P}_{\ell}({\cal V})$ is the set of all
vectors which have empirical distribution $P$, that is, 
\[
{\cal T}_{\ell}(P)=\text{\ensuremath{\left\{  y\in{\cal V}^{\otimes\ell}\colon P(v)=\frac{1}{\ell}\sum_{i=1}^{\ell}\I\{y(i)=v\}\text{ for all }v\in{\cal V}\right\} } ,}
\]
and the number of types is 
\[
\tau_{\ell}:=\left|{\cal P}_{\ell}({\cal V})\right|=\binom{\ell+|{\cal V}|-1}{|{\cal V}|-1}.
\]
Let us enumerate the set of $\ell$-types over ${\cal V}$ (empirical
distributions) as ${\cal P}_{\ell}({\cal V})=\{\hat{P}_{\ell}(1),\hat{P}_{\ell}(2),\ldots,\hat{P}_{\ell}(\tau_{\ell})\}$.
In the framework of Sec. \ref{sec:A-Unified-Online}, we may now consider
the action set ${\cal A}=[m]$. In this case, the action is a choice
of a type in ${\cal P}_{\ell}({\cal V})$. Given that the action is
$a\in[\tau_{\ell}]$ the rule $f$ determines that the distribution
over ${\cal Y}$ is uniform distribution over ${\cal T}_{\ell}(\hat{P}_{\ell}(a))$,
that is,
\[
Q(y)=\begin{cases}
\frac{1}{|{\cal T}_{\ell}(\hat{P}_{\ell}(a))|}, & y\in{\cal T}_{\ell}(\hat{P}_{\ell}(a))\\
0, & \text{otherwise}
\end{cases}.
\]
Codebooks drawn from such distributions are known as \emph{constant
composition codebooks} \cite[Ch. 9]{csiszar2011information}. In our
context, a uniform distribution over a type class can be thought of
as a ``pure action''. We next generalize the set of possible actions
to ``mixed actions'', in which the reconstruction distribution is
a \emph{mixtures} over types. In this case, the action set is ${\cal P}([\tau_{\ell}])$,
and an action can be written as $a\equiv(a(1),a(2),\ldots,a(\tau_{\ell}))$,
with weights $a(j)>0$, which sum to one, i.e., $\sum_{j=1}^{\tau_{\ell}}a(j)=1$.
The rule that determines the reconstruction distribution based on
the action $a$ is a mixture of uniform distributions over type classes.
That is, if $y\in{\cal T}_{\ell}(\hat{P}_{\ell}(j))$ for $j\in[\tau_{\ell}]$,
then 
\[
Q_{t}(y)=\frac{a(j)}{\left|{\cal T}_{\ell}(\hat{P}_{\ell}(j))\right|},
\]
that is, $a(j)$ is the probability that the reconstruction has type
$\hat{P}_{\ell}(j)$, and the reconstruction is chosen uniformly at
random within this type. 

\paragraph*{Equivalence of Memoryless Distributions and Mixture over Uniform
Type Distributions}

As mentioned, the optimal reconstruction distribution in the case
of a memoryless source is a memoryless distribution. We next shortly
outline its connection to mixture over uniform distributions over
types. Suppose that the $\ell$ symbols of $y$ are generated in an
memoryless manner, according to some distribution $P_{V}\in{\cal P}({\cal V})$,
that is $y(i)\sim P_{V}$, IID. Such a distribution results a distribution
over ${\cal Y}={\cal V}^{\otimes\ell}$, and thus a distribution in
${\cal P}([\tau_{\ell}])$ over the $\tau_{\ell}$ possible type classes
in ${\cal P}_{\ell}({\cal V})$. Specifically, for $P_{V}\in{\cal P}({\cal V})$
it holds that the probability of $\hat{P}_{\ell}(j)$ is given by
\begin{align}
a(j) & =\P\left[y\in{\cal T}_{\ell}(\hat{P}_{\ell}(j))\right]\\
 & =\left|{\cal T}_{\ell}(\hat{P}_{\ell}(j))\right|\cdot\exp\left[-\ell H(\hat{P}_{\ell}(j))-\ell D_{\text{KL}}(\hat{P}_{\ell}(j)\mid\mid P)\right]\\
 & =\exp\left[-\ell D_{\text{KL}}(\hat{P}_{\ell}(j)\mid\mid P)+O(\log\ell)]\right].\label{eq: probability of a typeclass under IID}
\end{align}
Hence, a memoryless distribution is a \emph{specific} mixture over
uniform distribution over types. We may then let ${\cal A}_{\ell}^{\circ}\subset{\cal P}([\tau_{\ell}])$
denote the set of possible mixture induced by memoryless distributions. 

We begin with the first version \cite{ZamirRose1997} of the NTS algorithms.
Suppose that the allowed set of reconstruction distributions includes
an arbitrary mixture of uniform distributions over type classes (which
also includes memoryless distributions). It then can be shown \cite{ZamirRose1997},
that in the asymptotic $\ell\to\infty$ regime, if $Q_{1}$ is a \emph{uniform}
mixture over the types then the type of $Y_{1}$ is $Q^{*}$, the
rate-distortion achieving reconstruction distribution. Therefore,
the pseudo-regret in this case is a constant which stagnates at $t=1$.
This agree with the well known fact that a \emph{uniform} mixture
over the types is a universal distribution \cite{ornstein1990universal,yu1993rate,zhang1997redundancy,mahmood2023lossy,merhav2023d,merhav2024lossy}.
The proof of this property heavily rely on asymptotic properties of
types, to wit, the fact that the number of types increases only polynomially
with $\ell$ \cite[Ch. 2]{csiszar2011information}, and it should
be noted that this universal mixture distribution is not memoryless. 

Specifically, the required rate, which is the minimal cost of (\ref{eq: optimal cost}),
is given by 
\[
R^{*}=R(P_{U},d)+O\left(\frac{\log\ell}{\ell}\right).
\]
This is the rate-distortion function plus a \emph{redundancy} term,
which vanishes with $\ell$. This cost can be achieved by drawing
reconstruction codewords randomly according to a proper \emph{universal
reconstruction distribution}. Nonetheless, such a scheme has a few
drawbacks. First, the universal distribution may be rather complex,
and accordingly, so is the codebook generation process. Second, the
sample $x$ might have intricate dependencies between its components,
and so the single-letter rate-distortion function $R(\hat{P}_{x},d)$
is overly pessimistic. While there are extensions of universal codes
to adapt to the actual rate-distortion function of the source, e.g.,
\cite{merhav2024lossy}, the redundancy rates are typically much slower.
Third, for continuous alphabets from source and reconstruction distributions
that are not easily parameterized, universal distributions are more
challenging to find, and the redundancy may be high.

The second version of NTS is based on memoryless distributions, as
the ones achieving the rate-distortion function (even at finite $\ell$),
and which are much simpler than arbitrary mixture distributions. NTS
\cite{ZamirRose1997,NTS_original,NTS_parametric,Zamir_NTS_cont,NTS_abstract_alphabet,NTS_ISIT,NTS_memory}
follows the sequential backward adaptation paradigm, and is a specific
\emph{expert-based} policy for determining the reconstruction distribution,
based on the history. The set of actions is ${\cal A}_{\ell}={\cal P}({\cal V})$,
and given an action $a=P_{V}$, the reconstruction distribution is
$Q=P_{V}^{\otimes\ell}$, that is, a memoryless distribution. At time
$t$, the policy only depends on the last observed reconstruction,
given by $Y_{t-1}$, that is, $\pi_{t}({\cal H}_{t})\equiv\pi_{t}(Y_{t-1})$.
The action is then determined as follows. Let the type $\hat{P}_{Y_{t-1}}\in{\cal P}({\cal V})$
of $Y_{t-1}=(Y_{t-1}(1),\ldots,Y_{t-1}(\ell))$ be the empirical distribution
\[
\hat{P}_{Y_{t-1}}(v)=\frac{1}{\ell}\sum_{i=1}^{\ell}\I\{Y_{t-1}(i)=v\},
\]
for $v\in{\cal V}$. The reconstruction distribution is then $Q_{t}=\hat{P}_{Y_{t-1}}^{\otimes\ell}\in{\cal P}({\cal V})$,
which is a memoryless distribution. Thus, perhaps $Q_{1}$, the actions
actually belong to the set of types ${\cal P}_{\ell}({\cal V})$,
which is a \emph{finite} set. This policy is myopic, in the sense
that the next action only depends on the type of the \emph{last} reconstructed
symbol $Y_{t-1}$. While this scheme is practically effective at any
reasonably large $\ell$, only asymptotic results are known on its
performance. It is known that for $\ell\to\infty$, the types of the
reconstructions generated by this method $\hat{P}_{Y_{t}}$ are the
ones that are also generated by the fixed-distortion Arimoto--Blahut
algorithm \cite{Blahut72,Arimoto72,Thomas_Cover} when initializing
from $P_{V}$ (which is equivalent to alternating minimization between
convex sets). Hence, it converges to the memoryless distribution achieving
the rate-distortion function $R(d,P_{U})$ as $t\to\infty$. In the
asymptotic case $\ell\to\infty$, the cost of using reconstruction
distribution $Q_{t}$ is the $R(P_{U},Q_{t},d)$, the rate-distortion
function assuming the reconstruction distribution is $Q_{t}$. The
pseudo-regret for the memoryless source $P_{U}$ is then \cite{Boukris_convergence}
\[
\pregret_{t}=\sum_{t=1}^{\infty}R(P_{U},Q_{t},d)-R(P_{U},d)\leq D_{\text{KL}}(Q_{V}^{*}\mid\mid Q_{1}),
\]
which is bounded, \emph{irrespective of the horizon}. Nonetheless,
a major issue with the mixtures defined by memoryless distributions
is that their weights vanish as $\ell\to\infty$ (\ref{eq: probability of a typeclass under IID}).
For mild values of $\ell$, the stochastic variability of the type
of the reconstruction $Y_{t}$ is rather large. NTS may not converge,
and bounds on the regret are unknown. Several numerical examples can
be found in \cite[Ch. 3.5]{elshafiy2022optimal}, e.g., Fig. 3.5 therein,
showing lack of stability even for $\ell=64$ after millions of iterations. 

The third version is \cite{NTS_ISIT}, which aims to further ameliorate
the asymptotic nature of NTS by averaging the types of $k>1$ reconstructions,
and then to update the reconstruction distribution once every $k$
time points. Specifically, an initial memoryless distribution $Q_{1}=P_{V}^{\otimes\ell}$
is chosen, and is used to generate the codebooks that will be used
in the first $k$ time points. Thus, $\pi_{2}({\cal H}_{2})=\pi_{3}({\cal H}_{3})=\cdots=\pi_{k}({\cal H}_{k})$
all result the same memoryless reconstruction $Q_{1}$. Then, at time
$k+1$, the empirical types of the reconstructions $Y_{1},Y_{2},Y_{k}$
are used to determine $Q_{k+1}$ as the memoryless distribution
\[
Q_{k+1}=\left(\frac{1}{k}\sum_{i=2}^{k+1}\hat{P}_{Y_{i-1}}\right)^{\otimes\ell}.
\]
The policy continues in this fashion, with sequences of $k$ rounds,
in which the $k$ types of the reconstructed symbols are averaged
to obtain a memoryless distribution. 

This scheme converges for any fixed $\ell$, when $k\to\infty$, and
then $n\to\infty$. Moreover, if the limit $\ell\to\infty$ is also
taken, then the recursion converges to the reconstruction distribution
achieving $R(P_{X},d)$. Nonetheless, when either $\ell$ or $k$
is small, NTS-based methods may not work well (except, perhaps, in
the memoryless very-low-distortion regime, where the observed types
$\hat{P}_{Y_{t}}$ are indicative of $P_{U}$, the memoryless distribution
generating $x$). When the distortion is not low, or in face of complex
statistical dependencies in $X_{t}$, MAB-based solutions are of interest. 

\section{MAB-based Lossy Compression Algorithms }

In what follows, we return to the general framework, and assume that
${\cal X}$ and ${\cal Y}$ are general alphabets, and that a set
of possible actions, determining reconstruction distributions, are
available. We focus on policies that only use the past actions and
observed costs, that is, the restricted history $\tilde{{\cal H}}_{t}:=(A_{1},B_{1},\ldots A_{t-1},B_{t-1})$,
to choose its next action. The incorporation of $\{Y_{1},\ldots,Y_{t-1}\}$
into this decision is more involved, and will be discussed in the
future research section. 

\subsection{Finite Action Sets}

In this section, we assume that the set of possible actions is finite
$|{\cal A}|=k$, and thus so is the set of possible reconstruction
distributions ${\cal Q}$. This set of actions is arbitrary, and should
be chosen based on prior knowledge of the MAB algorithm designer,
and leads to inductive bias. Accordingly, we analyze the regret with
respect to (w.r.t.) the best action in this set (which is a relative
measure of performance). For brevity, we enumerate them as $\{Q^{(i)}\}_{i\in[k]}\subset{\cal P}({\cal Y})$,
and for simplicity denote the set of actions by $[k]$. The MAB algorithm
we will utilize is the based on a lower confidence bound (LCB). Similarly
to the LCB algorithm with standard rewards \cite[Sec. 2.2]{bubeck2012regret}
\cite[Ch. 1]{slivkins2019introduction},\footnote{Equivalently, the algorithm is commonly known as the \emph{upper }confidence
bound (UCB), when the goal is to maximize reward and not minimize
cost.} the algorithm is based on a confidence interval estimate for the
expected cost of each of the $k$ arms. The specific challenge here
is that the expected cost is a rather complicated function of the
arm distribution. We thus next present a confidence interval on the
expected cost. 

Let us consider $x\in{\cal X}$ and a random codebook ${\cal C}$
drawn IID from a distribution $Q$. Then, since the codewords are
IID, it holds that 
\[
J(x,{\cal C})\sim\text{Geometric}(p(x,Q))
\]
for some unknown $d$-match probability parameter 
\[
p(x,Q):=\P_{Y\sim Q}\left[\rho(x,Y)\leq d\right].
\]
For MAB algorithms, UCB are constructed based on concentration inequalities
(e.g., for bounded or Bernoulli MAB problems, Hoeffding or Bernstein
inequalities are directly used \cite[Ch. 2]{bubeck2012regret,slivkins2019introduction}),
and so our goal is to derive a similar one for the average cost
\[
\tilde{b}(Q):=\E_{X\sim P_{X},{\cal C}\sim Q}\left[\log J(X,{\cal C})\right].
\]
The technical aspect is to derive a suitable confidence interval for
the cost $\tilde{b}(Q)$, which, as the logarithm of a geometric random
variable, follows a non-standard distribution. This is addressed in
the following proposition.
\begin{prop}
\label{prop: cost empirical estimation}Let a reconstruction distribution
$Q$, a distortion measure $\rho$, and $\delta\in(0,1)$ be given.
Assume that for some $\eta>0$ the minimal $d$-match probability
is lower bounded as $\min_{x\in{\cal X}}p(x,Q)>\eta$. Let $X_{t}\sim P_{X}$
IID, and let ${\cal C}_{t}\sim Q$ be a codebook with IID codewords
drawn from $Q$. There exists a numerical constant $c>0$ such that
with probability $1-\delta$
\[
\left|\frac{1}{n}\sum_{t=1}^{n}\log J(X_{t},{\cal C}_{t})-\tilde{b}(Q)\right|\leq\frac{c}{\eta}\sqrt{\frac{1}{n}\log\frac{1}{\delta}}.
\]
\end{prop}
\begin{IEEEproof}[Proof outline]
 $J(x,C_{t})$ is geometric random variable. We show that $\log J(x,{\cal C})$
is sub-exponential with Orlicz norm upper bounded by $1/\log(\frac{1}{1-p(x,Q)}).$
The same can be shown for $J(X_{t},{\cal C}_{t})$, which is a mixture
of geometric random variables, though with $p(x,Q)$ replaced by its
minimal possible value. The concentration of the empirical mean $\frac{1}{n}\sum_{t=1}^{n}\log J(X_{t},{\cal C}_{t})$
to the true mean $\tilde{b}(Q)$ then follows from Bernstein's inequality.
See proof in Appendix \ref{sec:Proof of empirical cost}. 
\end{IEEEproof}
The regret bound is inversely proportional to $\eta$, the minimal
$d$-match probability. In the high-distortion regime, $\eta=\min_{x\in{\cal X}}p(x,Q)$
is large, and the confidence interval shrinks. However, in case $X_{t}$
represents a length-$\ell$ vector, then this minimal probability
decreases with $\ell$, typically exponentially. Nonetheless, recent
work \cite{ryu2025improved} hints that this condition might be relaxed,
and this is left for future research. We also remark that slightly
modifying the scheme allows to control $\eta$. We can use an \emph{escape}
procedure whenever a $d$-match is not found with until some (large)
index. The encoder indicates this with a flag, and then sends a detailed
description of $X_{t}$, which has high, yet bounded, cost. In this
case the dependence in the bound on $\eta$ is replaced by the escape
probability.

\paragraph*{Operation of the algorithm}

A listing of the algorithm is in Appendix \ref{sec: LCB algorithm},
and here we just describe its operation. The encoder and decoder track
a confidence interval for the mean cost of each of the $k$ actions.
This confidence interval is computed from Prop. \ref{prop: cost empirical estimation},
using the costs incurred when using this reconstruction distribution
for encoding past symbols. 

For an action $a\in[k]$, let $N_{t}(a)$ be number of times that
action $a$ has been played up to round $t$, and let the $N_{t}(a)$
costs obtained by $a$ up to time $t$ by ${\cal B}_{t}(a)$. At time
$t$, the next action, $A_{t}$, determines a reconstruction distribution
$Q_{t}=Q^{(A_{t})}$, to generate the next codebook ${\cal C}_{t}$.
It is chosen as the action that minimizes the lower confidence interval.
A new sample $X_{t}$ from the source is compressed by ${\cal C}_{t}$,
and its cost $B_{t}=b(X_{t},{\cal C}_{t})$ (number of bits required
for compression) is added to the other costs in ${\cal B}_{t}(A_{t})$. 

The algorithm takes as input the distortion function $\rho$, the
required distortion level $d$, a sequence of lower confidence bounds
for the cost $\{\lcb(\cdot,\cdot,n)\}_{n\in\mathbb{N}_{+}}$, a parameter
$\alpha>2$, and a confidence function $\delta_{t}\equiv\delta_{t}(\alpha)$,
which determines the required reliability of the confidence interval
at any round $t$. 

Let the LCB function be
\[
\lcb((B_{1},\ldots,B_{n}),\delta,n)\equiv\frac{1}{n}\sum_{t=1}^{n}B_{t}-\frac{c}{\eta}\sqrt{\frac{1}{n}\log\frac{1}{\delta}},
\]
where $c>0$ is the constant from Prop. \ref{prop: cost empirical estimation}.
At each time point, the encoder and decoder optimistically choose
the next action as the one minimizing the LCB
\[
A_{t}\in\argmin_{a\in[k]}\lcb({\cal B}_{t}(a),\delta_{t},N_{t}(a))\}.
\]
The output of the algorithm is the chosen action $\{A_{t}\}_{t\in\mathbb{N}_{+}}$
at each round, which determines $\{N_{t}(a)\}_{a\in[k],\;t\in\mathbb{N}_{+}}$,
the number of times that each of the actions $a\in[k]$ have been
chosen up to time $t$. Let the optimal action be 
\[
a^{*}:=\argmin_{a\in[k]}\tilde{b}(Q^{(a)})
\]
 and let the gap of action $a$ be 
\[
\Delta(a):=\tilde{b}(Q^{(a)})-\tilde{b}(Q^{(a^{*})}),
\]
which is the excess cost of action $a$ over the optimal action $a^{*}$.
Given the algorithm's output, the pseudo-regret at round $t$ is given
by $\sum_{a\in[k]:\Delta(a)>0}N_{t}(a)\Delta(a)$, whose expected
value is $\pregret_{t}$, as in (\ref{eq: pseudo regret}). 

We then have the following bound on the pseudo regret. 
\begin{thm}
\label{thm: UCB regret}Assume that the MAB algorithm is run with
arbitrary ${\cal Q}=\{Q^{(a)}\}_{a\in[k]}$ of gaps $\{\Delta(a)\}_{a\in[k]}$,
and $\delta_{t}=t^{-\alpha}$ for some $\alpha>2$. Then, the pseudo-regret
is bounded as 
\begin{equation}
\pregret_{t}\leq\sum_{a\in[k]:\Delta(a)>0}\left[\frac{4c^{2}\alpha}{\eta^{2}\Delta(a)}\log(t)+\frac{2(\alpha-1)}{\alpha-2}\cdot\Delta(a)\right].\label{eq: regret of bias}
\end{equation}
\end{thm}
\begin{IEEEproof}
The lower confidence bound is such that at time $t$, for a gap $\Delta>0$,
and $\delta_{t}=t^{-\alpha}$ the number of samples $n$ required
to make the deviation term less than $\Delta/2$ is $n\geq\frac{4c^{2}\alpha}{\eta^{2}\Delta^{2}}\log(t)$.
The result then follows from standard analysis of such algorithms,
e.g. \cite[Proof of Th. 2.1]{bubeck2012regret}.
\end{IEEEproof}
Using the standard routine, we next obtain from Theorem \ref{thm: UCB regret}
a gap-independent bound. Taking, for simplicity, a constant $\alpha$,
and focusing on the first term which is the dominating one (as it
increases with $t$ and blows-up with $\Delta(a)$), then $\Delta_{\text{min}}:=\min_{a\in[k]\backslash\{a^{*}\}}\Delta(a)$
the pseudo-regret is upper bounded as $O(k\frac{\log(t)}{\eta^{2}\Delta_{\text{min}}}).$
The pseudo-regret is also upper bounded by $\Delta_{\text{max}}t$,
where $\Delta_{\text{max}}:=\max_{a\in[k]}\Delta(a)=\lambda\Delta_{\text{min}}$,
where $\lambda>0$ is the ratio between the maximal and minimal gap,
and so
\[
\pregret_{t}=O\left(k\frac{\log(t)}{\eta^{2}\Delta_{\text{min}}}+\lambda\Delta_{\text{min}}t\right).
\]
Maximizing the above right-hand side leads to the worst-case gap $\Delta_{\text{min}}=\Theta(\sqrt{k\frac{\log(t)}{\eta^{2}\lambda t}})$,
and thus to:
\begin{cor}
\label{cor: LCB worst case}For a fixed $\lambda\geq0$ it holds that
\[
\pregret_{t}=O\left(\sqrt{\frac{k}{\lambda\eta^{2}}\cdot t\log(t)}\right).
\]
\end{cor}
Beyond the usual dependence on the number of arms, the bound is inversely
proportional to $\eta=\min_{x\in{\cal X}}p(x,Q)$. When the distortion
increases, this probability increases too, and the regret bound is
improved.

\subsection{Lipschitz Bandits: Exploiting Action Similarity }

In this section, we exemplify that whenever there are plausible structural
assumption on the actions, the pseudo-regret can be sub-linear even
for infinite action sets. We consider the case of a finite alphabet
$|{\cal Y}|<\infty$. In this case each $Q\in{\cal P}({\cal Y})$
is a possible reconstruction distribution, and thus the number of
possible actions is uncountable. Naturally, however, if $Q_{1}$ and
$Q_{2}$ are close in some sense, then we may expect that their mean
cost is similar. Indeed, Jensen's inequality readily implies that
mean cost of $Q$ satisfies 
\[
\tilde{b}(Q):=\E_{X\sim P_{X},{\cal C}\sim Q}\left[\log J(X,{\cal C})\right]\leq\E_{X\sim P_{X}}\left[\log\frac{1}{p(X,Q)}\right].
\]
Thus, we expect that $\tilde{b}(Q_{1})-\tilde{b}(Q_{2})$ should be
related to $\log\frac{p(x,Q_{1})}{p(x,Q_{2})}$. We first note that:
\begin{claim}
It holds that 
\[
\mu(Q_{1},Q_{2}):=\max_{x\in{\cal X}}\left|\log\frac{p(x,Q_{1})}{p(x,Q_{2})}\right|
\]
is a metric over ${\cal P}({\cal Y})$: It is non-negative and symmetric,
and it is simple to prove it satisfies a triangle inequality. 
\end{claim}
Since $p(x,Q)=\sum_{y\in{\cal Y}\colon\rho(x,y)\leq d}Q(y)$, this
metric depends on the distortion level $d$. We next establish that
$Q\mapsto b(Q)$ is a Lipschitz-continuous function over ${\cal P}({\cal Y})$,
w.r.t. the metric $\mu(Q_{1},Q_{2})$.
\begin{prop}
\label{prop: Lipschitz continuity}It holds that $|\tilde{b}(Q_{1})-\tilde{b}(Q_{2})|\leq\mu(Q_{1},Q_{2})$
for any $Q_{1},Q_{2}\in{\cal P}({\cal Y})$. 
\end{prop}
See proof in Appendix \ref{sec:Proof Lipscihtz continuity}. Thus
we may \emph{cover} the space of possible distributions ${\cal P}({\cal Y})$
by a net $\{Q_{*}^{(a)}\}_{a\in[k]}$ (a finite covering set) such
that for any $Q\in{\cal P}({\cal Y})$ there exists $a\in[k]$ such
that $\mu(Q,Q_{*}^{(a)})\leq\epsilon$. In the pseudo-regret analysis
we assumed that $\min_{x\in{\cal X}}p(x,Q)>\eta>0$, so we slightly
restrict the set of possible distributions to the set
\[
{\cal P}^{(\eta)}({\cal Y}):=\left\{ Q\in{\cal P}({\cal Y})\colon Q(y)\geq\eta\text{ for all }y\in{\cal Y}\right\} .
\]
 The next lemma provides the size of a proper net for ${\cal P}^{(\eta)}({\cal Y})$. 
\begin{lem}
\label{lem: covering net size}Let $\epsilon>0$ and $\eta\in(0,1)$
be given. There exists a set $\{Q_{*}^{(a)}\}_{a\in[k]}$ of cardinality
\[
k_{*}(\epsilon)=\left(\frac{1}{\eta\epsilon}\sqrt{\max_{x\in{\cal X}}\sum_{y\in{\cal Y}}\I\{\rho(x,y)\leq d\}}\right)^{|{\cal Y}|-1}
\]
 such that for any $Q\in{\cal P}^{(\eta)}({\cal Y})$ there exists
$a\in[k]$ such that $|\tilde{b}(Q)-\tilde{b}(Q_{*}^{(a)})|\leq\epsilon$.
\end{lem}
See proof in Appendix \ref{sec: proof of covering net size lemma}.
The complexity term $\max_{x\in{\cal X}}\sum_{y\in{\cal Y}}\I\{\rho(x,y)\leq d\}$
is the maximal number of $d$-matches that any $x\in{\cal X}$ can
have, and it increases with the distortion level $d$.

Now, we may use the LCB algorithm from the previous section, with
the net $\{Q_{*}^{(a)}\}_{a\in[k]}$ as the $k$ possible set of actions.
Optimizing the value of $k$ results the following upper bound on
the pseudo-regret:
\begin{thm}
\label{thm: UCB Lipschitz regret}Let $\eta\in(0,1)$ be given, and
consider a MAB problem with reconstruction distribution set of ${\cal P}^{(\eta)}({\cal Y})$.
Assume that the LCB algorithm is run with ${\cal Q}=\{Q_{*}^{(a)}\}_{a\in[k]}$
being a chosen net of ${\cal P}^{(\eta)}({\cal Y})$. Then, at time
horizon $T$ is 
\begin{equation}
\pregret_{T}=O\left(\left(\Gamma T^{|{\cal Y}|}\log T\right)^{1/(|{\cal Y}|+1)}\right)\label{eq: regret of Lipchitz algorithm}
\end{equation}
where 
\[
\Gamma:=\frac{1}{\lambda\eta^{|{\cal Y}|+1}}\left(\max_{x\in{\cal X}}\sum_{y\in{\cal Y}}\I\{\rho(x,y)\leq d\}\right)^{(|{\cal Y}|-1)/2}.
\]
\end{thm}
Note that the bound is horizon-dependent, but can be made horizon-free
using the usual doubling-trick.
\begin{IEEEproof}
We follow the analysis of Lipschitz bandits \cite[Ch. 4]{slivkins2019introduction}.
Let $\epsilon>0$ be given. Using Lemma \ref{lem: covering net size},
we construct a net of $k_{*}(\epsilon)$ points. At time-horizon $T$,
the approximation error is upper bounded by $\epsilon T$. Utilizing
Corollary (\ref{cor: LCB worst case}) with $k=k_{*}(\epsilon)$,
the pseudo-regret is upper bounded as 
\[
\pregret_{T}=O\left(\frac{\sqrt{\Gamma T\log(T)}}{\epsilon^{(|{\cal Y}|-1)/2}}+\epsilon T\right).
\]
Optimizing the discretization level as $\epsilon=((\Gamma\log T)/T)^{1/(|{\cal Y}|+1)}$
directly leads to the bound. 
\end{IEEEproof}
Evidently, in the general case, while the bound of Theorem \ref{thm: UCB Lipschitz regret}
is sub-linear $\tilde{O}(T^{|{\cal Y}|/(|{\cal Y}|+1)})$, and is
interesting in the regime of low $|{\cal Y}|$. However, it severely
suffers from the curse-of-dimensionality, even for moderate values
of $|{\cal Y}|$. 

\section{Summary and Future Directions}

We proposed a unified framework for codebook adaptation in universal
sequential lossy compression, and inspected both classic NTS solutions
and new MAB-based approaches. The NTS is mostly suitable under strong
assumptions on the source distribution (e.g., memoryless), and its
adaptation relies on asymptotic properties on the typical behavior
of past reconstructions. Theoretical guarantees assure the optimality
of this method, though mostly in the asymptotic regime, and show that
the regret is bounded for all $t$ by a constant. 

By contrast, our MAB algorithm is more robust, as it makes no assumptions
on the intra-symbol structure within $x$ (thus works for any $\ell$),
and adapts its codebook based only on the incurred cost. Theoretical
guarantees assure a sub-linear regret, though one that grows without
bound, as $O(\sqrt{kt\log t})$ when the $k$-actions are pre-specified,
and at a much higher rate, as $\tilde{O}(T^{|{\cal Y}|/(|{\cal Y}|+1)})$,
when covering the simplex of possible distributions over a finite
reconstruction alphabet. 

Can we enjoy both worlds? Namely, for a small-to-medium block length
$\ell$, can we combine the MAB robustness with the NTS structural
advantages? The challenge is to incorporate the reconstruction history
$\{Y_{i}\}_{i\in[t-1]}$ into the policy. Consider a reconstruction
distribution that is based on a uniform mixture of \emph{two} types.
Suppose that at some $t$, the first $d$-match is from the first
type, with an index codeword $j=j_{1}+j_{2}$, where $j_{1}$ (resp.
$j_{2}$) is the number of codewords with the first (resp. second)
type, up to index $j$ (we expect that $j_{1}\approx j_{2}\approx j/2$).
The NTS will use the first type in the next round, but the actual
feedback of the decoder for this action is that the first type generates
a $d$-match after $j_{1}$ attempts, whereas the second type requires
more than $j_{2}$ attempts. This can be used to update some quality
measure function for \emph{both} types, even for the one that did
not result the $d$-match. This idea can be generalized to a mixture
of multiple types, and results an interesting exploration-exploitation
trade-off, between using the best type thus far, versus instantaneously
exploring multiple types at once, though with a lower quality feedback
on their quality. These promising directions are left for future investigation.

\section*{Acknowledgments}

The authors thank Kenneth Rose and Alon Peled-Cohen for multiple interesting
discussions regarding this research.

\appendices{\numberwithin{equation}{section}}

\section{Proof of Prop. \ref{prop: cost empirical estimation} \label{sec:Proof of empirical cost}}
\begin{lem}
\label{lem: empirical estimation of log geometric mixture}Let $X\sim P_{X}$
be a random variable over ${\cal X}$, which has a density $\nu(x)$
w.r.t. some base measure $\mu(x)$, that is, $\nu(x)=\frac{\d P_{X}}{\d\mu(x)}$.
Further let $\eta\in(0,1)$ be given, and let $p(x)\equiv p(x,Q)\in(\eta,1)$
be a parameter for each possible realization of $X$. Let $U$ be
a mixture of geometric distributions with parameter $p(x)$ and mixture
probability $P_{X}$, that is, for $u\in\mathbb{N}_{+}$ 
\[
\P[U=u]=\int(1-p(x))^{u-1}p(x)\cdot\nu(x)\d\mu.
\]
Let $U_{i}\sim U$ for $i\in[n]$ be IID. Then, for $\delta\in(0,1/2)$
there exists a numerical constant $c>0$ such that with probability
$1-\delta$
\[
\left|\frac{1}{n}\sum_{i=1}^{n}\log U_{i}-\E[\log U]\right|\leq\frac{c}{\eta}\sqrt{\frac{1}{n}\log\frac{1}{\delta}}.
\]
\end{lem}
\begin{IEEEproof}
Let us first evaluate the tail of $\log U$. It holds that 
\begin{align}
\P\left[\log U\geq t\right] & =\P\left[U\geq e^{t}\right]\\
 & =\int\P\left[U\geq e^{t}\mid X=x\right]\nu(x)\d\mu\\
 & \leq\int\left(1-p(x)\right){}^{e^{t}}\nu(x)\d\mu\\
 & \leq\int\left(1-\eta\right){}^{e^{t}}\nu(x)\d\mu\\
 & \leq(1-\eta)^{1+t}\\
 & =(1-\eta)\cdot\exp\left[-\log\left(\frac{1}{1-\eta}\right)\cdot t\right].
\end{align}
So, the centering property \cite[Ex. 2.7.10]{vershynin2018high} implies
that 
\[
\left\Vert \log U-\E[\log U]\right\Vert _{\psi_{1}}\lesssim\frac{1}{\log\left(\frac{1}{1-\eta}\right)}.
\]
Hence, by Bernstein's inequality \cite[Th. 2.8.1]{vershynin2018high},
there exists a numerical constant $c>0$ such that 
\begin{align}
 & \P\left[\frac{1}{n}\sum_{i=1}^{n}\log U_{i}-\E[\log U]\geq t\right]\nonumber \\
 & \leq2\exp\left[-c\cdot n\left(\log^{2}\left(\frac{1}{1-\eta}\right)t^{2}\right)\wedge\left(\log\left(\frac{1}{1-\eta}\right)t\right)\right],
\end{align}
and for $t\leq\frac{1}{\log\left(\frac{1}{1-\eta}\right)}$
\[
\P\left[\frac{1}{n}\sum_{i=1}^{n}\log U_{i}-\E[\log U]\geq t\right]\leq2\exp\left[-c\cdot n\log^{2}\left(\frac{1}{1-\eta}\right)t^{2}\right],
\]
which means that with probability $1-\delta$
\[
\left|\frac{1}{n}\sum_{i=1}^{n}\log U_{i}-\E[\log U]\right|\leq\sqrt{\frac{1}{n}\cdot\frac{\frac{1}{c_{0}}\log\frac{2}{\delta}}{\log^{2}\left(\frac{1}{1-\eta}\right)}.}
\]
There exists a numerical constant $c_{1}>0$ so that 
\[
\log^{2}\left(\frac{1}{1-\eta}\right)\geq c_{1}\eta^{2}
\]
and then 
\[
\left|\frac{1}{n}\sum_{i=1}^{n}\log U_{i}-\E[\log U]\right|\leq\frac{1}{\eta}\sqrt{\frac{1}{n}\cdot\frac{1}{c_{0}c_{1}}\log\frac{2}{\delta}}
\]
with probability $1-\delta$. 
\end{IEEEproof}

\section{A Detailed Listing of the LCB Algorithm \label{sec: LCB algorithm}}

\begin{algorithm}
\begin{algorithmic}[1]

\Procedure{LCB lossy compression}{$k$, $\{Q^{(i)}\}_{i\in[k]}$$\rho(\cdot,\cdot)$,
$d$ ,$\lcb(\cdot,\cdot,n)$, $\{\delta_{t}\}_{t\in\mathbb{N}_{+}}$
}

\State Encoder and decoder initialize ${\cal B}_{1}(a)=\emptyset$
and $N_{1}(a)=0$ for all $a\in[k]$

\Comment{ The cost observation set of each action is empty at round
$t=0$}

\For{ $t=1,2,\ldots$}

\State  \textbf{Encoder and decoder find} 
\[
A_{t}\in\argmin_{a\in[k]}\lcb({\cal B}_{t}(a),N_{t}(a),\delta_{t})\}
\]

\State  \textbf{Encoder and decoder generate a random codebook ${\cal C}_{t}=(Y_{t}(1),Y_{t}(2),\ldots)\sim Q^{(A_{t})}$}

\State  \textbf{Encoder observes $X_{t}$, and finds 
\[
J(X_{t},{\cal C}_{t})=\inf_{j\in\mathbb{N}_{+}}\left\{ j\colon\rho\left(X_{t},Y_{t}(j)\right)\leq d\right\} 
\]
}

\State  \textbf{Encoder sends $J(X_{t},{\cal C}_{t})$ to the decoder
using $B_{t}=b(X_{t},{\cal C}_{t})$ bits}

\State  \textbf{Decoder receives $J(X_{t},{\cal C}_{t})$ and reconstructs
$Y_{t}={\cal C}_{t}(J(X_{t},{\cal C}_{t}))$}

\State  \textbf{Encoder and decoder update} 
\begin{align}
{\cal B}_{t+1}(A_{t}) & \leftarrow({\cal B}_{t}(A_{t}),B_{t})\\
N_{t+1}(A_{t}) & \leftarrow N_{t}(A_{t})+1
\end{align}
\Comment{ The cost observation of the chosen action is added to the
set of observations}and for $a\in[k]\backslash A_{t}$
\begin{align}
{\cal B}_{t+1}(A_{t}) & \leftarrow{\cal B}_{t}(A_{t})\\
N_{t+1}(A_{t}) & \leftarrow N_{t}(A_{t})
\end{align}

\Comment{The observation set of others arms is unchanged}

\EndFor

\State \textbf{Return} $\{N_{t}(a)\}_{a\in[K],\;t\in\mathbb{N}_{+}}$

\Comment{The number of times each arm $a\in[k]$ have been played
up to each round $t\in\mathbb{N}_{+}$}

\EndProcedure

\end{algorithmic}

\caption{An LCB algorithm for backward adaptive lossy compression \label{alg:A-UCB-general}}
\end{algorithm}

\section{Proof of Prop. \ref{prop: Lipschitz continuity} \label{sec:Proof Lipscihtz continuity}}
\begin{IEEEproof}[Proof of Prop. \ref{prop: Lipschitz continuity}]
To evaluate the mean value, we use the identity
\[
\E[\log U]=\int_{0}^{\infty}\left[e^{-t}-\E[e^{-tU}]\right]\frac{\d t}{t}
\]
and use for $t\ge0$ the MGF of a geometric random variable 
\[
\E[e^{-tU}]=\frac{pe^{-t}}{1-(1-p)e^{-t}}
\]
for which it holds that 
\begin{align}
e^{-t}-\E[e^{-tU}] & =e^{-t}-\frac{pe^{-t}}{1-(1-p)e^{-t}}\\
 & =e^{-t}\cdot\frac{(1-p)(1-e^{-t})}{1-(1-p)e^{-t}}
\end{align}
to obtain 
\begin{align*}
\E[\log U] & =\int_{0}^{\infty}e^{-t}\cdot\frac{(1-p)(1-e^{-t})}{1-(1-p)e^{-t}}\cdot\frac{\d t}{t}.
\end{align*}
Now, for $Q_{1},Q_{2}$ and a given $x$, let $p_{1}\equiv p(x,Q_{1})$,
$p_{2}\equiv p(x,Q_{2})$, and assume that $p_{1}<p_{2}$. Then, 
\begin{align}
 & \left|\overline{b}(x,Q_{1})-\overline{b}(x,Q_{1})\right|\nonumber \\
 & =\left|\E_{{\cal C}\sim Q_{1}}\left[\log J(X,{\cal C})\right]-\E_{{\cal C}\sim Q_{2}}\left[\log J(X,{\cal C})\right]\right|\\
 & \leq\int_{0}^{\infty}\frac{e^{-t}(1-e^{-t})}{t}\cdot\left|\frac{(1-p_{1})}{1-(1-p_{1})e^{-t}}-\frac{(1-p_{2})}{1-(1-p_{2})e^{-t}}\right|\cdot\d t\\
 & \leq\int_{0}^{\infty}e^{-t}\cdot\left|\frac{(1-p_{1})}{1-(1-p_{1})e^{-t}}-\frac{(1-p_{2})}{1-(1-p_{2})e^{-t}}\right|\cdot\d t\\
 & \leq|r_{1}-r_{2}|\cdot\int_{0}^{\infty}e^{-t}\d t\\
 & =\left|\log p_{1}-\log p_{2}\right|,
\end{align}
where in the first inequality we used the integral formula, and in
the second inequality we used $|1-e^{-t}|\leq t$ for $t\geq0$. For
the third inequality, we let $r_{1}=\log p_{1}\in\mathbb{R}_{-}$
and $r_{2}=\log p_{2}\in\mathbb{R}_{-}$, and consider the function
\[
f(r):=\frac{(1-e^{r})}{1-(1-e^{r})e^{-t}}
\]
for which 
\begin{align}
f'(r) & =\frac{-e^{r}\left[1-(1-e^{r})e^{-t}\right]-(1-e^{r})e^{r-t}}{\left[1-(1-e^{r})e^{-t}\right]^{2}}\\
 & =\frac{-e^{r}+e^{r-t}-e^{2r-t}-e^{r-t}+e^{2r-t}}{\left[1-(1-e^{r})e^{-t}\right]^{2}}\\
 & =\frac{-e^{r}}{\left[1-(1-e^{r})e^{-t}\right]^{2}}.
\end{align}
For $t>0$ and $r<0$ it holds that 
\[
1-e^{r/2}\geq(1-e^{r})e^{-t}
\]
and so
\[
\left[1-(1-e^{r})e^{-t}\right]^{2}\le e^{r}
\]
hence $|f'(r)|\leq1$ and $r\mapsto f(r)$ is Lipschitz. We thus conclude
that
\begin{align}
 & \left|\tilde{b}(Q_{1})-\tilde{b}(Q_{2})\right|\\
 & =\left|\E_{X\sim P_{X}}\left[\E_{{\cal C}\sim Q_{1}}\left(\log J(X,{\cal C})\right)-\E_{{\cal C}\sim Q_{1}}\left(\log J(X,{\cal C})\right)\right]\right|\\
 & \leq\left|\E_{X\sim P_{X}}\left[\log\frac{p(X,Q_{1})}{p(X,Q_{2})}\right]\right|\\
 & =\left|\sum_{x\in{\cal X}}P_{X}(x)\log\frac{p(X,Q_{1})}{p(X,Q_{2})}\right|\\
 & \leq\max_{x\in{\cal X}}\left|\log\frac{p(x,Q_{1})}{p(x,Q_{2})}\right|,
\end{align}
as claimed.
\end{IEEEproof}

\section{Proof of Lemma \ref{lem: covering net size} \label{sec: proof of covering net size lemma}}
\begin{IEEEproof}[Proof of Lemma \ref{lem: covering net size}]
Note that we may write
\[
p(x,Q)=\P_{Y\sim Q}\left[\rho(x,Y)\leq d\right]=\sum_{y\in{\cal Y}}Q(y)\I\left\{ \rho(x,y)\leq d\right\} .
\]
Let us consider the simplex over ${\cal Y}$ as embedded in $\mathbb{R}^{|{\cal Y}|}$.
If we take $Q\in\mathbb{R}^{|{\cal Y}|}$ and also consider $\I\{\rho(x,y)\leq d\}$
as a vector $v_{d,x}\in\{0,1\}^{|{\cal Y}|}\subset\mathbb{R}^{|{\cal Y}|}$
then we may write
\[
p(x,Q)=\langle Q,v_{d,x}\rangle
\]
(with the usual inner product). Then, from Prop. \ref{prop: Lipschitz continuity},
a sufficient condition for $|\tilde{b}(Q_{1})-\tilde{b}(Q_{2})|\leq\epsilon$
is 
\[
\left|\log\langle Q_{1},v_{d,x}\rangle-\log\langle Q_{2},v_{d,x}\rangle\right|\leq\epsilon
\]
for all $x\in{\cal X}$. Note that 
\[
\log\langle Q,v_{d,x}\rangle=\log\sum_{y}Q(y)\I\left\{ \rho(x,y)\leq d\right\} \leq\log1=0
\]
and 
\[
\log\langle Q,v_{d,x}\rangle\geq\log\min_{x}\sum_{y}Q(y)\I\left\{ \rho(x,y)\leq d\right\} \geq\log\eta.
\]
since we assumed that for any $x\in{\cal X}$ there exists $y\in{\cal Y}$
such that $\rho(x,y)\leq d$. Let us cover the $(|{\cal Y}|-1)$ dimensional
cube with a uniform grid ${\cal G}$ of resolution $\beta$. This
can be achieved with $\beta^{-(|{\cal Y}|-1)}$ points. So, for any
$Q$ there exists $Q_{*}\in{\cal G}$ so that $\|Q-Q_{*}\|_{\infty}\leq\beta$.
Thus, since 
\[
\eta\leq\langle Q,v_{d,x}\rangle\leq1
\]
and since $t\mapsto\log t$ is $(1/\eta)$-Lipschitz on $[\eta,1]$
it holds from the Cauchy--Schwarz inequality that 
\begin{align}
 & \left|\log\langle Q,v_{d,x}\rangle-\log\langle Q_{k}^{*},v_{d,x}\rangle\right|\nonumber \\
 & \leq\frac{1}{\eta}\cdot\left|\langle Q-Q_{k}^{*},v_{d,x}\rangle\right|\\
 & \leq\frac{1}{\eta}\cdot\sqrt{\max_{x\in{\cal X}}\sum_{y\in{\cal Y}}v_{d,x}(y)}\cdot\beta
\end{align}
for all $x\in{\cal X}$. 
\end{IEEEproof}
\bibliographystyle{ieeetr}
\bibliography{RD_seq}

\end{document}